\documentstyle[12pt]{article}
\begin{document}
\title{Integrability in the mesoscopic dynamics}
\author{Artur Sowa\\
109 Snowcrest Trail,
Durham, NC 27707 \\
ArturSowa@mesoscopia.com  }
\date{}
\maketitle
\newtheorem{th}{Theorem}[section]
\newtheorem{cor}{Corollary}[section]
\newtheorem{pr}{Proposition}[section]
\newtheorem{df}{Definition}[section]

The Mesoscopic Mechanics (MeM), as introduced in \cite{MeM}, is relevant to the 
electron gas confined to two spatial dimensions.
It predicts a special way of \emph{collective} response of \emph{correlated} electrons to the external magnetic field. The dynamic variable of this theory is a finite-dimensional operator, which is required  to satisfy the mesoscopic Schr\"{o}dinger equation, cf.  (\ref{MesoSchr}) below.

In this article, we describe general solutions of the mesoscopic Schr\"{o}dinger equation.
Our approach is specific to the problem at hand.
It relies on the unique structure of the equation and makes no reference to any other techniques, with the exception of the geometry of unitary groups.
In conclusion, a surprising fact comes to light. Namely, the mesoscopic dynamics ``filters" through the (microscopic) Schr\"odinger dynamics as the latter turns out to be a clearly separable part, in fact an autonomous factor, of the evolution. This is a desirable result also from the physical standpoint. 

\section{A brief description of contents and results}

The mesoscopic Schr\"odinger equation describes evolution of an operator (denoted $K$) via a nonlinear equation. In order to motivate the reader let me point out that the interpretation of this operator is somewhat similar to that of the wavefunctions (of the regular Schr\"odinger equation). Informally speaking, the pair of operators $K$ and $K^*$ may be interpreted as essentially being square roots of a density matrix, (cf. \cite{MeM}), although the issue is delicate due to non-commutativity. This should be viewed as comparable to the fact that modulus-square of a wavefunction  represents a probability distribution. I have proposed the mesoscopic  Schr\"odinger equation  as a model for correlated evolution of an n-tuple of electrons, which is relevant to the galvanomagnetic properties of the so-called correlated materials.  

In this article, I present analysis of the evolution of the system when the single particle Hamiltonian, which is an ingredient in the equations, does not receive any feedback from the dynamic variable $K$.  The last section of this article, Section \ref{sec4}, provides a brief summary of the physical interpretation of the mesoscopic equation. In particular, it should explain why there is incentive also to consider the case when such a feedback would exist. This problem is not addressed in the present article. However, as regards the case limited to the $K$-independent Hamiltonian $H$, the problem is essentially resolved. 

Section \ref{sec1} is meant to introduce the system, and to review some basic properties, display simple special solutions. Also, it is pointed out that the mesoscopic equation has the structure of a Hamiltonian system. However, I emphasize, no further use is made of the so-called canonical formalism.  Next, in Section \ref{sec2}, the equation is solved in the case when the domain and image of the operator $K$ are fixed finite-dimensional spaces. In fact, it is shown that the evolution of this nonlinear system with time-dependent Hamiltonian can be represented in a certain way via a pair of curves on the unitary group.  Finally, we consider the case when the domain and the image of operator $K$ are {\em a priori} allowed to evolve in an ambient Hilbert space. In this case, the single-particle Hamiltonian is densely defined on the Hilbert space. Solutions of the mesoscopic equation in such a broad setting are described in Section \ref{sec3}. In particular, a uniqueness property is shown. Its proof takes advantage of an exceptional structure of the mesoscopic equation and it could not, it seems, be deduced from any general principles. Also, it is shown that the evolution is ``driven" by an n-tuple of Schr\"odinger particles. This is important from the physical stand-point as obtaining any other type of carriers might be problematic from the standpoint of physical interpretation.  In addition, we note that evolution entails a phase factor which explicitly depends on the history of the magnetic energy density $B^2(t)$. Thus, in the fixed-domain and the moving-domain cases alike, solutions are represented by means of simpler factors corresponding to certain linear problems. We emphasize that this remains valid even when the constituents of the equation are time dependent. This is what is meant by \emph{integrability in the mesoscopic dynamics}.

Let me emphasize that while considerations in Section \ref{sec2} are related to the ordinary differential equations, those of Section \ref{sec3} deal with partial differential equations. However, I believe, the context of operator equations with evolving domains may be quite new. It enables one to capture essentially new phenomena that cannot be discussed on grounds of the PDE setting alone.     
 
\section{The mesoscopic Schr\"odinger equation}
\label{sec1}

Let $F$ be a finite-dimensional complex vector space equipped with a Hermitian scalar product. Let $t$ denote the time variable, and let
\[
H(t): F\rightarrow F,\]
be a predetermined family of positive definite self-adjoint operators, which we will refer to as the Hamiltonian. In addition,  let $B=B(t)$ be a predetermined function of time, which we will refer to as the magnetic induction. We will require throughout this article that both $B$ and $H$ depend on the time variable smoothly. This is a technical assumption, which will ensure local existence and uniqueness of solutions of certain  dynamical systems that we will encounter along the way. 
Introduce the dynamic operator variable 
\begin{equation}
K(t):F\rightarrow G.
\label{defK}
\end{equation}
We assume that $K$ has a null kernel, $\ker K = \{0\},$ while the target space $G = \mbox{Im}(K(t))$ is an arbitrary but fixed complex linear space also equipped with a Hermitian scalar product. (In fact, we will consider a more general situation in Section \ref{sec3}.)
Throughout this article our attention is focused on the mesoscopic Schr\"{o}dinger equation
\begin{equation}
\label{MesoSchr}
i\hbar\dot{K} = -KH  -B^2 (K^*)^{-1},
\end{equation}
where the $*$ denotes Hermitian conjugation. Note that the nonlinearity is of a homogeneous type but develops a singularity as $\det{K}\rightarrow 0$, which {\em a priori} may be an intimidating factor as one attempts to solve the equation. 

Let us recall here that the manifold of invertible linear transformations, say, from $F$ to $G$, is equipped with a natural Hermitian metric given by
\[
\langle L|N\rangle = \mbox{trace}\left(L N^* \right).
\] 
Here $L$ and $N$ denote two arbitrary tangent vectors which, let it be emphasized, represent arbitrary linear transformations from $F$ to $G$. Furthermore, the Hermitian structure induces a compatible Riemannian structure 
\[
\langle L,N\rangle = \Re \left\{\mbox{trace}\left(L N^* \right)\right\},
\]
as well as a symplectic form
\[
\omega (L,N) =  \Im\left\{\mbox{trace} \left(L N^* \right)\right\}.
\]
With this understood, let us point out that the evolution equation (\ref{MesoSchr}) is tied to the following {\em total} Hamiltonian 
\begin{equation}
\Xi (K) = \mbox{trace} \left(K H K^* \right) + B^2\log \det\left(KK^* \right).
\label{Ksi}
\end{equation}
Indeed, a calculation shows that the differential of $\Xi$ is given by
\begin{equation}
\label{derivative}
\begin{array}{ll}
d\Xi _K[L]=\left. \frac{d}{d\varepsilon}\right| _{\varepsilon =0}\Xi (K+\varepsilon L) &= \Re \left\{\mbox{trace}\left(\left(K H +B^2(K^*)^{-1}\right) L^* \right) \right\} \\
 &= \langle K H +B^2(K^*)^{-1},L\rangle,
\end{array}
\end{equation}
Furthermore, since 
\[
\Re\left\{\mbox{trace} \left(A B^* \right)\right\} = \Im\left\{\mbox{trace} \left(iA B^* \right)\right\},
\] 
equation (\ref{derivative}) can be re-interpreted in the form 
\[
d\Xi _K[L]= \omega \left(i\left(K H +B^2(K^*)^{-1}\right), L\right).
\]
This means precisely that (\ref{MesoSchr}) is the \emph{Hamiltonian flow} (cf. \cite{Arnold}) induced by the total Hamiltonian $\Xi$ and the symplectic structure $\omega$.
While it is good to bring this theme to the reader's attention, it will not be explicitly essential to the discussion in this article.

Formula (\ref{derivative}) indicates that the critical points of the Hamiltonian $\Xi$, subject to the constraint $\mbox{trace} \left(K K^* \right) = \mbox{const}$, satisfy the Euler-Lagrange equation
\begin{equation}
KH  +B^2 (K^*)^{-1} = \nu K.
\label{Euler-Lagrange}
\end{equation}
This equation implies
\[
K^*K( \nu - H)  = B^2\mbox{Id}.
\]
In addition, since $K^*K>0$, the equation can be satisfied only if the real scalar $\nu$ dominates all the eigenvalues of $H$. In conclusion, all solutions of (\ref{Euler-Lagrange}) are of the form
\begin{equation}
K_\nu=U\frac{B}{(\nu-H)^{1/2}}, 
\label{solutionK}
\end{equation}
where $U:F\rightarrow G$ is an arbitrary unitary operator, and $\nu$ is arbitrary as long as it dominates $H$. The critical points are interesting in their own right, cf. \cite{MeM}. In addition, they play a special role in the time-dependent problem (\ref{MesoSchr}). Indeed, assume for a while that  the Hamiltonian $H$ is time-independent and diagonalized by vectors $|\psi _n\rangle \in F$, so that
\begin{equation}
H|\psi _n\rangle = E_n |\psi _n\rangle 
\label{Schrodprime}
\end{equation}
for a collection of positive eigenvalues $E_n$. Apparently, the simplest solutions of equation (\ref{MesoSchr}) are  of the form 
\begin{equation}
K = \sum a_n(t)|\psi _n\rangle\langle \psi_n|,
\label{async}
\end{equation}
where $a_n=r_ne^{i\varphi_n}$. Substituting this into (\ref{MesoSchr}) one readily obtains
\begin{equation}
r_n = r_{n,0}\quad\mbox{and}\quad \varphi_n = \frac{1}{\hbar}\left(E_n + \frac{B^2} {r_{n,0}^2}\right)t +\varphi_{n,0}.
\label{a_n}
\end{equation}
It is interesting to note that when all $a_n$'s are correlated, i.e. oscillate with the common frequency, say, $\nu = E_n + B^2/r_{n}^2$ for all $n$, then 
\[
r_n = \frac{\pm B}{(\nu - E_n)^{1/2}}.
\]
Therefore, these special solutions conform with (\ref{solutionK}), and so they represent critical points of the Hamiltonian $\Xi$.

\section{General fixed-domain solution of the mesoscopic equation}
\label{sec2}

In this section, it will be shown that equation (\ref{MesoSchr}) can be reduced to a system of simpler equations, even when  $H$ and $B$, i.e. the constituents of the equation, are time dependent. As we set out to solve the equation, the first useful artifice is to use polar representation of the operator. Namely, let 
\begin{equation}
\label{polar}
K=RU,
\end{equation}
where $R=R^*$ is positive definite, and  $U^{-1}=U^*$, i.e. $U$ is unitary. It ought to be emphasized that here the matrix $U$ stands on the right, which is in contrast to the situation in (\ref{solutionK}). For a given $K$, its polar representation is determined by setting
\[
R= \sqrt{KK^*}:G\rightarrow G\mbox{, and } U=R^{-1}K : F\rightarrow G. 
\]
A direct calculation shows that $U$ selected in this way is unitary. It is well known that with the requirement of positive definiteness of $R$ the polar decomposition is unique. Next, observe that when $K$ satisfies equation (\ref{MesoSchr}), then
\[
\begin{array}{rllll}
\hbar\frac{d}{ dt}(R^2) = \hbar\frac{d}{dt}(KK^*) =& \\
= & \hbar\dot{K}K^* +\hbar K\dot{K^*}  \\
= & -i\left(-KH  -B^2 (K^*)^{-1}\right)K^*+iK(-HK^*  -B^2 K^{-1} )  \\
= & iKHK^*+iB^2-iKHK^*-iB^2 \\
=& 0.
\end{array}
\]
It ought to be emphasized again that the calculation remains valid whether or not the Hamiltonian and the magnetic field depend on the time variable. Let us now set $K(0)= K_0$. We have  
\[
R^2 = K_0K_0^*.
\]
There is only one positive definite, self-adjoint $R$ satisfying this condition. Since $K_0K_0^*$ is positive definite and self-adjoint, it can be diagonalized in a certain basis so that 
\[
 K_0K_0^* = \mbox{diag} \left[\lambda _1^2,\lambda _2^2, \ldots ,\lambda _N^2 \right],
\]
and, in the same basis,
\[
R = \mbox{diag} \left[|\lambda _1 |,|\lambda _2 | , \ldots ,|\lambda _N | \right].
\] 
$R$ does not depend on time. In summary, we obtain  
\begin{cor} 
Evolution prescribed by equation (\ref{MesoSchr}) is constrained to the submanifold 
\[
M_R=\left\{ K: KK^* = R^2 \right\}. 
\]
 All coordinate functions of the matrix $KK^*$ are integrals of motion. As is easily seen, $M_R$ is diffeomorphic with the unitary group and has half the dimension of the phase space.
\end{cor}
A similar calculation as above shows that 
\begin{equation}
\label{derksi}
\frac{d}{ dt}\Xi (K(t)) = \mbox{trace}\left(K \dot{H}K^* \right) + 2B\dot{B}\log \det R^2 .
\end{equation}
In particular, as the system evolves, change in the entropy part of the total Hamiltonian only depends on $B(t)$. If $H$ does not depend on time, then $\mbox{trace}\left(K HK^* \right)$ is an additional integral of motion.

We now continue to discuss solutions of (\ref{MesoSchr}). First, denote
\[
U(0)  = R^{-1}K_0= U_0.
\]
Next, substitute $K$ in its polar representation into equation (\ref{MesoSchr}) to obtain
\[
i\hbar R\dot {U}
=  -RUH  -B^2 R^{-1}U  
\]
Multiplying the equation by $R^{-1}$ one further obtains
\begin{equation}
\label{MesoSchrUn}
i\hbar\dot{U} = -UH  - H_BU, 
\end{equation}
where 
\begin{equation}
H_B = B^2R^{-2}= B^2(K_0K_0^*)^{-1}.
\label{hb}
\end{equation}
In this way, evolution of the unitary part is determined by the predetermined constituents $H$ and $B$ as well as the initial condition $K_0$. In fact, it may be more practical for some purposes to represent equation (\ref{MesoSchrUn}) in the form
\begin{equation}
\label{MesoSchrUnG}
U^*\dot{U} =  \frac{i}{\hbar}(H  + U^*H_BU).
\end{equation}
The left-hand side represents a vector tangent to the trajectory, shifted to the group unit. The right hand side, driving the evolution, represents an element in the Lie algebra of skew-Hermitian operators. Indeed, not only $H$ but also $H_B$, and hence also $U^*H_BU$ are Hermitian operators.  Suppose at first that $H$ and $B$ are all frozen in time. Since the unitary group is compact and the group multiplication is smooth, the right-hand side of (\ref{MesoSchrUn}) defines a Lipschitz continuous vector field on the unitary group. In particular, solutions of (\ref{MesoSchrUn}) are uniquely determined (via a choice of the initial condition) and exist for all time. In fact, in this case the solution may be written in the form of a power series
\begin{equation}
\label{series}
\begin{array}{ll}
U(t)  =  U_0 & + \frac{it}{\hbar}(U_0H+H_BU_0)-\frac{t^2}{ 2!\hbar ^2}(U_0H^2+2H_BU_0H +H_B^2U_0) \\ \\
 & -  i\frac{t^3}{3!\hbar ^3}(U_0H^3+3H_BU_0H^2 +3H_B^2U_0H+H_B^3U_0) + \ldots, 
\end{array}
\end{equation}
 Since all operators are finite dimensional, the series converges absolutely. A straightforward calculation shows that $U(t)$ satisfies (\ref{MesoSchrUn}). 

We proceed to resolving the case when $H$ and $B$ are allowed to vary in time smoothly.  First, represent $U$ as a product of two unitary matrices, i.e.
\[
U(t)=V(t)W(t).
\]
Equation (\ref{MesoSchrUn}) yields
\begin{equation}
\label{MesoSchrUndouble}
i\hbar\dot{V}W +i\hbar V\dot{W} = -VWH  - H_BVW. 
\end{equation}
Secondly, multiply the equation by $V^*$ on the left and by $W^*$ on the right. This leads to
\begin{equation}
\label{MesoSchrsplit}
i\hbar V^* \dot{V} +i\hbar\dot{W}W^* = -WHW^*  - V^*H_BV. 
\end{equation}
Now, the two factors have been separated. Indeed, ask that $W$ and $V$ satisfy the following two separate equations 
\begin{equation}
\label{MesoSchrright}
i\hbar\dot{W} = -WH, 
\end{equation}
and
\begin{equation}
\label{MesoSchrleft}
i\hbar \dot{V} = - H_BV.
\end{equation}
In the case of time-varying $H$ and $H_B$ only local existence of solutions of (\ref{MesoSchrUn}), (\ref{MesoSchrright}), and  (\ref{MesoSchrleft}) is guarantied, but the uniqueness property is still retained. It follows that if $U(0)=V(0)W(0)$, then $U(t)=V(t)W(t)$ for all $t$. 
Note that $H_B$ depends on time only via $B$, and due to Hermicity, it can be written in a certain basis as
\[
H_B(t) = B^2(t)\mbox{diag} \left[\lambda _1^{-2},\lambda _2^{-2}, \ldots ,\lambda _N^{-2} \right].
\]
Thus, the solution of (\ref{MesoSchrleft}) can be represented in the same basis in the form 
\[
\begin{array}{ll}
V(t) &= \exp\left(\frac{i}{\hbar}\int_{0}^{t}B^2(t')(K_0K_0^*)^{-1}dt' \right) \\
\\
&=
\mbox{diag} \left[\exp\left(\frac{i}{\hbar}\lambda _1^{-2}\int_{0}^{t}B^2(t')dt'\right),
 \ldots ,
\exp\left(\frac{i}{\hbar}\lambda _N^{-2}\int_{0}^{t}B^2(t')dt' \right)\right].
\end{array}
\]
Here, we have selected the initial condition $V(0)=Id$. This needs to be compensated by the appropriate choice of the second initial condition, namely $W(0)=U_0$. As it turns out, we have essentially reduced equation (\ref{MesoSchr}) to a pair of simpler, well-understood equations.  Let us summarize the results.
\begin{th}
Consider the mesoscopic Schr\"{o}dinger equation (\ref{MesoSchr}) with smooth constituents $H=H(t)$  and $B=B(t)$. The solution $K=K(t)$ satisfying the initial condition 
\[
K(0) = K_0=RU_0
\]
is a uniquely defined smooth operator-valued function of time. Furthermore, the solution admits representation in the from 
\begin{equation}
K(t)=\sqrt{K_0K_0^*}\exp\left(\frac{i}{\hbar}\int_{0}^{t}B^2(t')(K_0K_0^*)^{-1}dt' \right)W(t),
\end{equation}
where $W$ satisfies
\[
i\hbar\dot{W} = -WH(t), \qquad W(0) = U_0. 
\] 
When $H$ and $B$ are time-independent the solution exists for all time, while in general it is only guarantied to exist locally. 
\label{theor1}
\end{th}
Of course, if both $H_B$ and $H$ are time-independent, then $V(t)$ and $W(t)$ represent two geodesics of the bi-invariant metric on the unitary group, e.g. cf. \cite{Milnor}. They can also be represented as power series. One can perform multiplication of the two series and grouping of the terms to see that the product is equivalent to the series in equation (\ref{series}). It is worthwhile to mention that when $H$ depends on time, $W(t)$ can still be represented in terms of the time-ordered exponential, cf. \cite{Grein} p. 219.

It is worthwhile to substitute the solution of (\ref{MesoSchr}) in the form specified in Theorem \ref{theor1} into formula (\ref{derksi}). A calculation involving the property that $\mbox{trace}\left(AB \right)=\mbox{trace}\left(BA \right)$ shows that the following holds.
\begin{cor} In the notation of Theorem \ref{theor1}, we have
\begin{equation}
\label{derksipost}
\frac{d}{ dt}\Xi (K) = \mbox{trace}\left( W^*R^2W\dot{H} \right) + \frac{d}{ dt}(B^2)\log \det R^2 .
\end{equation}
Since $R$ is fixed in time, the magnetic (entropy) part of the energy only depends on magnetic induction during the evolution. Recall that evolution of  $W$ only depends on $H$, and so the electronic part of the energy is only affected by the electronic constituent. (Of course, $H$ could depend on $B$ via, say, Landau quantization.)
\end{cor}

 I would also like to highlight the fact that  we have made many arbitrary choices when solving equation (\ref{MesoSchr}). Naturally, we have made those choices so as to simplify the discussion. In spite of that, uniqueness of solutions guaranties that the result is general. One of the very conspicuous arbitrary choices was declaring time-independent operator $R$. We need not impose the condition of positive definiteness of $R$. If that condition is dropped and when $K_0K_0^*$ has degenerate eigenvalues, one can select a time-varying $R$ satisfying the constraint $R^2 = K_0K_0^*$.  Subsequently, one would redefine the auxiliary Hamiltonian $H_B$ by setting $H_B=B^2R^{-2} + i\hbar R^{-1}\dot{R}$. Naturally, this would also redefine $V=V(t)$ and in the end yield the same product $RV$ as the calculation based on the time-independent $R$.

Let us look back at the findings in this section. Recall that equation (\ref{MesoSchr}) has a strong yet homogeneous nonlinearity. In fact, one might argue it is quadratic in nature. Our approach was to exploit the underlying group structure. Specifically, the polar decomposition of the dynamic variable allowed us to reduce the nonlinear initial value problem to a pair of linear-type evolution problems.  Naturally, Theorem \ref{theor1} implicitly makes a reference to Quantum Mechanics (via the operator $W$). In fact, the inter-connectedness of Quantum Mechanics and the Mesoscopic Mechanics will come to sharper focus in the next section.

\section{General evolving-domain solution of the mesoscopic equation}
\label{sec3}

In the previous sections we have worked under the assumption that the domain and image of the operator $K$ defined in (\ref{defK}) are frozen in time. However, this assumption is neither necessary nor natural in the context of the mesoscopic  equation (\ref{MesoSchr}). Indeed, it is natural to consider a more general setting when {\em a priori} both the domain and the image of operator $K$ are allowed to evolve, i.e.
\begin{equation}
K_{FG}(t):F(t)\rightarrow G(t)
\label{defKt}
\end{equation}
Here, it is understood that 
\[
F(t) \subset \mbox{\textsf{H}}_1, \mbox{ and } G(t) \subset\mbox{\textsf{H}}_2
\]
are finite-dimensional subspaces in two (possibly different) infinite-dimensional (separable) Hilbert spaces. 
In particular, the spaces $F(t)$ and $G(t)$ all inherit the Hermitian structure from the ambient Hilbert spaces.  Furthermore, in this context,  consider the Hamiltonian 
\[
H(t):\mbox{\emph{D}}\rightarrow\mbox{\textsf{H}}_1,
\] 
which is well defined on a (fixed in time) dense linear subspace 
\[
\mbox{\emph{D}}\subseteq\mbox{\textsf{H}}_1.
\]
$H(t)$ are also (formally) self-adjoint, i.e. 
\begin{equation}
\label{Hself}
\langle \varphi |H(t)\psi \rangle=\langle H(t)\varphi |\psi \rangle \mbox{ for all }\varphi , \psi\in\mbox{\emph{D}}.
\end{equation}
For a reason that will soon become clear we require {\em a priori} that 
\begin{equation}
\label{Findom}
F(t) \subset\mbox{\emph{D}}
\end{equation}
throughout the evolution.
Finally, let us emphasize that the particular realization of the Hilbert spaces $\mbox{\textsf{H}}_1$ and $\mbox{\textsf{H}}_2$ as well as the Hamiltonian $H(t)$ will remain implicit throughout our discussion as it is of no consequence to the conclusions we wish to draw.  

It is now clear how to interpret the mesoscopic equation (\ref{MesoSchr}) within this framework. Specifically, one needs to extend the operators $K_{FG}(t)$ through zero to the orthogonal complement of $F(t)$. Also, all operators $K_{FG}^*$ and $K_{FG}^{-1}$ need to be extended in an analogous way. Introduce the following \emph{shorthand} notation
\begin{equation}
\label{formal1}
K= K_{FG}\oplus 0_{F^\bot},
\end{equation}
and
\begin{equation}
\label{formal2}
K^{-1}= K_{FG}^{-1}\oplus 0_{G^\bot}.
\end{equation}
One checks directly that
\begin{equation}
\label{formal3}
K^*= K_{FG}^*\oplus 0_{G^\bot},
\end{equation}
and, moreover,  
\begin{equation}
\label{formal4}
(K^{-1})^* = (K^*)^{-1}.
\end{equation}
The shorthand notation seems intuitive and self-explanatory, and should not be confusing. We will refer to time-dependent families of operators of this type as the moving-domain operators. This terminology makes no reference to the `moving image' as indeed, we will show that the image remains fixed for solutions of the mesoscopic equation, cf. Theorem \ref{theor2}.

\begin{df}
\label{definemove}
We say that a moving-domain operator $K(t)$ as above is a local solution of (\ref{MesoSchr}) if for all $\psi\in\mbox{\emph{D}}$, equation 
\begin{equation}
\label{MesoSchrwk}
\left(i\hbar\dot{K} + KH  + B^2 (K^*)^{-1}\right)|\psi\rangle =0
\end{equation}
holds for all $t$ within a certain interval, say, $t\in [0, \varepsilon )$. Of course, we write
\[
i\hbar\dot{K} = -KH  -B^2 (K^*)^{-1}.
\]
\end{df}

The first goal is to show that the mesoscopic equation (\ref{MesoSchr}) has the uniqueness property even in this setting. In order to demonstrate this, the approach developed in the previous section will be exploited again.
First, observe that, in view of (\ref{formal2}), (\ref{formal3}), and (\ref{formal4}), equation (\ref{MesoSchrwk}) implies that  for all $e\in\mbox{\textsf{H}}_2$ and all $\psi\in\mbox{\emph{D}}$
\[
\langle \left(-i\hbar\dot{K^*} + H^*K^*  +B^2 K^{-1}\right)e|\psi\rangle =0.
\]
Hence, the {\em a priori} assumptions (\ref{Hself}) and (\ref{Findom}) allow us to conclude that
\begin{equation}
\label{MesoSchrstar}
i\hbar\dot{K^*} = H^*K^*  +B^2 K^{-1}.
\end{equation}
Of course, the latter equation is understood in the ordinary sense. (If this may at first seem puzzling, let us point out that $K^*$ sends all vectors from $\mbox{\textsf{H}}_2$ into $F(t) \subset\mbox{\emph{D}}$. Therefore the equation can be `evaluated' on all vectors from $\mbox{\textsf{H}}_2$, and so it is expected to hold therein as, in fact, it does.) Next, observe that for an arbitrary $e\in \mbox{\textsf{H}}_2$,
\[
\begin{array}{rllll}
i\hbar\frac{d}{dt}(KK^*)e =& \\
= & i\hbar\dot{K}K^*e +i\hbar K\dot{K^*}e  \\
= & \left(-KH  -B^2 (K^*)^{-1}\right)K^*e+K(H^*K^* + B^2 K^{-1} )e  \\
= & -KHK^*e -B^2e +KH^*K^*e + B^2e \\
=& 0.
\end{array}
\]
Here, the last equality is justified by the {\em a priori} assumptions (\ref{Hself}) and (\ref{Findom}). In particular, it follows that since $G(t)$ is the image of $KK^*$, it cannot evolve in time, i.e.
\[
G(t) = G(0).
\]
Therefore, it is possible to represent solutions in the polar decomposition with the self-adjoint and positive definite radial part $R:G\rightarrow G$, which is time independent. Now, suppose contrary to our expectation that (\ref{MesoSchr}) admits two {\em a priori} different moving-domain solutions on the interval $t\in [0,\varepsilon )$, say,
\[
 K_0(t) = R U_0(t),
\]
and
\[
 K_1(t) = R U_1(t),
\]
while initially 
\[
U_0(0) = U_1(0).
\]
Here,
\[
U_0(t), U_1(t): F(t)\rightarrow G,
\]
and the conventional extension to the whole space is understood implicitly. 
A direct calculation shows that 
\[
i\hbar\dot{U}_{0,1} = -U_{0,1}H  - H_BU_{0,1}, 
\]
where $H_B = B^2R^{-2}$.  Now, observe
\[
\begin{array}{rllll}
i\hbar\frac{d}{dt}\left(U_0U_1^*\right) = &
 -U_0HU_1^*  - H_BU_0U_1^* +U_0H^*U_1^* +U_0U_1^*H_B \\
 = &  - H_BU_0U_1^*+U_0U_1^*H_B,
\end{array}
\]
where, again, cancellation of two terms is justified by (\ref{Hself})  and (\ref{Findom}). At this stage, the extension of operator $U_1^*$ to the whole of $\mbox{\textsf{H}}_2$ plays no role. In fact, we can view $Y= U_0U_1^* :G\rightarrow G$ as being the finite dimensional unitary operator satisfying
\[
i\hbar\frac{d}{dt}\left(Y\right) =   -H_BY +YH_B.
\]
This is an equation of the type considered in Section \ref{sec2}, cf. equation (\ref{MesoSchrUn}).  
We already know it has the uniqueness property. Therefore, $Y(t) =Id$ is the unique solution of this equation with the initial condition $Y(0) =Id$. Thus, $U_0(t)U_1(t)^*=Id$, i.e. $U_0(t)=U_1(t)$ for all $t$. In summary, we have
\begin{th}
\label{uniqueness}
If the mesoscopic equation (\ref{MesoSchr}) in the broader moving-domain interpretation (cf. Definition \ref{definemove}) has a local solution $K(t)$ in the interval, say, $t\in [0, \varepsilon)$, then such a solution is uniquely defined by the initial condition $K=K(0)$.
\end{th}
The uniqueness property of (\ref{MesoSchr}) in such a broad Hilbert-space interpretation is a beautiful fact, indeed. Its proof relies on the inherent structure of the equation.    

Having established uniqueness of solutions we are empowered to find out the general form of solutions. Indeed, all we need to do is display a solution general enough to satisfy an arbitrary initial condition. Then, the uniqueness property will assure that no other solutions have been overlooked. This being the case, it would suffice to guess solutions, as long as they would be general enough. In what follows, it is shown how the general form of solutions can be deduced.

In order to shed some light on the nature of moving-domain solutions, consider first a simpler case when $F(t)= \mbox{span}\left\{|\psi (t)\rangle\right\}$, and $G(t)= \mbox{span}\left\{|\varphi (t)\rangle\right\}$, i.e. both spaces remain one-dimensional. Let operator $K$ be represented in the form 
\begin{equation}
\label{k1d}
K(t) = a(t) |\varphi (t)\rangle\langle\psi (t)|,
\end{equation}
where $a$ is a complex-valued function of time. Substituting, we find that equation (\ref{MesoSchr}) is translated into the following relation 
\[
i\hbar \left(\dot{a} |\varphi \rangle\langle\psi | + a |\dot{\varphi }\rangle\langle\psi | + a |\varphi \rangle\langle\dot{\psi }| \right) =
- a |\varphi \rangle\langle\psi |H - \frac{B^2}{a^*}|\varphi \rangle\langle\psi |.
\]
This latter equation is consistent if and only if there exist complex-valued functions of time $c_1(t)$ and $c_2(t)$ such that 
\begin{equation}
\label{phi}
i\hbar |\dot{\varphi }\rangle = c_1(t) |\varphi \rangle,
\end{equation}
\begin{equation}
\label{psi}
i\hbar \langle\dot{\psi }| = \langle\psi | \left(c_2(t)-H\right),
\end{equation}
and hence
\begin{equation}
\label{a}
i\hbar\dot{a} = -(c_1(t)+c_2(t))a-\frac{B^2(t)}{a^*}.
\end{equation}
Of course, the general solution of (\ref{phi}) is given by
\begin{equation}
\label{phi1}
|{\varphi (t)}\rangle = \exp {\left(-\frac{i}{\hbar}\int_{0}^{t}c_1(t')dt'\right)} |\varphi (0) \rangle.
\end{equation}
Next, introduce a new variable $\langle\psi '(t)|$, which is defined as follows
\begin{equation}
\label{psiprime}
\langle\psi '(t)| = \exp {\left(\frac{i}{\hbar}\int_{0}^{t}c_2(t')dt'\right)} \langle\psi (t)|.
\end{equation}
The benefit of this is that
\begin{equation}
\label{psi1}
i\hbar \langle\dot{\psi '}| = -\langle\psi '| H.
\end{equation}
Redefine also $a$ by setting
\begin{equation}
\label{aprime}
a' = a\exp {\left(-\frac{i}{\hbar}\int_{0}^{t}(c_1(t')+c_2(t'))dt'\right)}.
\end{equation}
Observe that in particular $K$ can now be re-written in the form
\begin{equation}
\label{k1dprime}
K(t) = a'(t) |\varphi (0)\rangle\langle\psi '(t)|.
\end{equation}
Moreover, substituting (\ref{aprime}) in (\ref{a}) yields
\begin{equation}
\label{a1}
i\hbar\dot{a'} = -\frac{B^2(t)}{a'^*}.
\end{equation}
Furthermore, setting $a'=r\exp {\left(i\Phi\right)}$ leads to
\[
\left( i\hbar\dot{r}-\hbar r\dot{\Phi}\right)\exp {\left(i\Phi\right)} = -\frac{B^2}{r}\exp {\left(i\Phi\right)}.
\]
Now, since the exponential factor cancels, the real and the imaginary parts of the equation can be separated. In conclusion 
\[
r=r_0 = \mbox{const},\quad  \Phi = \frac{1}{\hbar r_0^2}\int_{0}^{t}B^2(t')dt' +\Phi _0.
\]
We summarize the result as
\begin{pr}
Consider operators $K(t):F(t)\rightarrow G(t)$, where $F(t)$ and $G(t)$ are one-dimensional spaces for all $t$. Equation (\ref{MesoSchr}) admits solutions in this form if and only if the following two conditions hold: 
\begin{enumerate}
\item
\label{cond1}
The target space $G(t)= G(0) = \mbox{span}\left\{|\varphi (0)\rangle\right\}$ remains frozen in time.
\item
\label{cond2}
There is a vector $\psi '\in\mbox{\textsf{H}}_1$ satisfying the one-particle Shr\"odinger equation
\begin{equation}
\label{k1dsch}
i\hbar \frac{d}{dt}\langle\psi '(t)| = -\langle\psi '(t)| H,
\end{equation}
which spans the domain spaces, i.e.
\[
F(t)=\mbox{span}\left\{|\psi '(t)\rangle\right\}.
\]
\end{enumerate}
When both conditions \ref{cond1} and \ref{cond2} are satisfied, then the general solution of (\ref{k1dsch}) admits representation in the form
\begin{equation}
\label{k1dsol}
K(t) = r_0e^{i\Phi _0}\exp{\left(\frac{i}{\hbar r_0^2}\int_{0}^{t}B^2(t')dt'\right)} |\varphi (0)\rangle\langle\psi '(t)|,
\end{equation}
where $r_0$ and $\Phi _0$ are arbitrary real numbers.
\label{prop1} 
\end{pr}

\noindent

I would like to emphasize that in particular the problem of \emph{existence} of solutions of the nonlinear equation (\ref{MesoSchr}) has been reduced to the existence  property of the \emph{linear} Schr\"odinger equation (\ref{k1dsch}). Naturally, the existence result and other properties of the latter equation are well known, e.g. cf. \cite{Vlad}. Moreover, in view of this result, even the notion of \emph{regularity} of operator solutions of (\ref{MesoSchr}) acquires a clear meaning.

Next, let us consider the general case of $N$-dimensional domain and image spaces, which we will refer to as the $N\times N$-dimensional case. First, let
\[F(t)= \mbox{span}\left\{{|\psi _n (t)\rangle: n = 1\ldots N}\right\},
\]
 and 
\[
 G(t)=\mbox{span}\left\{{|\varphi _m(t)\rangle: m = 1\ldots N}\right\}.
\]
Furthermore, let $A$ be a complex matrix
\[
A(t) = [a_{m n}(t)]_{m, n = 1\ldots N}, \quad a_{m n} = [A]_{m n}.
\] 
Let the dynamic variable be represented in the form
\begin{equation}
\label{kNd}
K(t) = \sum a_{m n}(t) |\varphi _m (t)\rangle\langle\psi _n(t)|.
\end{equation}
(Summation is always carried out over repeated indices.)
Observe that in particular 
\begin{equation}
\label{kNdstarinv}
\left(K(t)^*\right)^{-1} = \sum [\left(A(t)^*\right)^{-1}]_{m n} |\varphi _m (t)\rangle\langle\psi _n(t)|.
\end{equation}
Initially, some progress is achieved by exploiting analogy with the  $1\times 1$-dimensional case. Indeed, observe that, in the $N\times N$-dimensional case, the mesoscopic Schr\"odinger equation (\ref{MesoSchr}) is translated into the following relation
\begin{equation}
\label{relat}
\begin{array}{ll}
&i\hbar\sum\left(\dot{a_{m n}}|\varphi _m \rangle\langle\psi _n| + a_{m n}|\dot{\varphi _m} \rangle\langle\psi _n| + a_{m n}|\varphi _m \rangle\langle\dot{\psi _n}|\right)=\\ \\
&-\sum a_{m n}(t) |\varphi _m (t)\rangle\langle\psi _n(t)|H - B^2\sum[ \left(A(t)^*\right)^{-1} ]_{m n} |\varphi _m (t)\rangle\langle\psi _n(t)|.
\end{array}
\end{equation}
Just as we have seen it before,
also here a simple linear consistency check will help draw far-reaching conclusions. First, observe that for the equation to hold there must exist complex functions of time $c'_{k m}(t)$ and $c''_{n l}(t)$ such that
\begin{equation}
\label{phimat}
i\hbar |\dot{\varphi _m}\rangle = \sum c'_{k m}(t) |\varphi _k \rangle \mbox { for all } m
\end{equation}
and
\begin{equation}
\label{psimat}
i\hbar \langle\dot{\psi _n}| + \langle\psi _n| H = \sum c''_{n l} \langle\psi _l| \mbox { for all } n.
\end{equation}
Secondly, introduce matrices 
\[
C'(t) = [c'_{k m}(t)]_{k, m  = 1\ldots N},\quad C''(t) = [c''_{n l}(t)]_{n, l = 1\ldots N}.
\]
Substituting (\ref{phimat}) and (\ref{psimat}) into (\ref{relat}), one obtains
\begin{equation}
\label{amat}
i\hbar\dot{A} +C'(t)A+AC''(t) = -B^2(t)\left(A^*\right)^{-1}.
\end{equation}
Note that
operator $K(t)$ as in (\ref{kNd}) satisfies (\ref{MesoSchr}) if and only if the three conditions (\ref{phimat}), (\ref{psimat}), and (\ref{amat}) are satisfied by the $|\varphi _k (t)\rangle $'s,  $\langle\psi _n (t)|$'s and the $A(t)$. In order to draw further conclusions, 
one ought to make the following observations. First,  one may require without loss of generality that both bases $|\varphi _m(t)\rangle$ and $\langle\psi _n|$ remain unitary during the evolution. Indeed, the operator $K(t)$ can be described in arbitrary bases of $F(t)$ and $G(t)$. Now, suppose the two bases are unitary, say, at  $t=0$. Equations (\ref{phimat}) and (\ref{psimat}) imply that the bases will remain unitary for all time if and only if  
\begin{equation}
\label{Chermit}
C'(t)^* = C'(t), \mbox{ and } C''(t)^* = C''(t),
\end{equation}
i.e. if these matrices are Hermitian. 
With this understood, denote by $\Gamma ' (t)$ and $\Gamma '' (t)$ the uniquely defined unitary matrices, which solve the two initial value problems:
\[
i\hbar\dot{\Gamma '} = C'(t)\Gamma ',\qquad \Gamma ' (0) = \mbox{Id},
\]
and
\[
i\hbar\dot{\Gamma ''} = \Gamma ''C''(t), \qquad \Gamma '' (0) = \mbox{Id}.
\]
In particular,  
\begin{equation}
\label{phimat1}
|\varphi _m (t)\rangle = \sum [\Gamma '(t)]_{km} |\varphi _k (0)\rangle \quad\mbox { for all } m.
\end{equation}
Next, define a new unitary collection of vectors $\langle\psi' _n(t)|$ as follows
\begin{equation}
\label{psimat1}
\langle\psi' _n(t)| = \sum [\Gamma ''(t)]_{nl} \langle\psi _l (t)|\quad \mbox { for all } n.
\end{equation}
Naturally, the collection $\langle\psi' _n(t)|$ provides new unitary bases for spaces $F(t)$. It has been selected in such a way as to simplify equation (\ref{psimat}). Indeed, a straightforward calculation shows that
\begin{equation}
\label{psimatschrod}
i\hbar \frac{d}{dt}\langle\psi' _n| = - \langle\psi' _n| H \quad \mbox{ for all } n.
\end{equation}
Furthermore, set
\begin{equation}
\label{amatprimeevol}
A'(t) = \Gamma '(t) A(t) \Gamma ''(t).
\end{equation}
and observe that (\ref{amat}) implies 
\begin{equation}
\label{amatprime}
i\hbar\dot{A'} = -B^2(t)\left(A'^*\right)^{-1}.
\end{equation}
We have already learned how to solve equations as this one in Section \ref{sec2}. Indeed, applying Theorem \ref{theor1} (with $H=0$) one obtains
\begin{equation}
\label{Atime}
A'(t)=\sqrt{A'(0)A'(0)^*}
\exp\left(\frac{i}{\hbar}\int_{0}^{t}B^2(t')(A'(0)A'(0)^*)^{-1}dt'\right).
\end{equation}
In fact, the initial conditions imply $A'(0) = A(0)$. 
Finally, observe that in view of (\ref{amatprimeevol}), (\ref{phimat1}), and (\ref{psimat1})
\[\begin{array}{lll}
K(t) & =  \sum [A]_{m n}|\varphi _m (t)\rangle\langle\psi _n(t)| \\ \\
& = \sum [\Gamma '(t)^{-1} A' \Gamma ''(t)^{-1}]_{m n}|\varphi _m (t)\rangle\langle\psi _n(t)| \\ \\
& = \sum [A']_{m n}|\varphi _m (0)\rangle\langle\psi '_n(t)|.
\end{array}
\]
In the end, one ought to substitute (\ref{Atime}) into the expression above. In summary, we have the following result

\begin{th}
Consider an operator $K(t):F(t)\rightarrow G(t)$, where $F(t)$ and $G(t)$ are evolving $N$-dimensional spaces. For $K(t)$ to satisfy the mesoscopic Schr\"odinger equation (\ref{MesoSchr}) in the sense of Definition \ref{definemove}, it is necessary and sufficient that the following two conditions be satisfied: 
\begin{enumerate}
\item
\label{cond1N}
The target space 
\[
G(t)=G(0) =\mbox{span}\left\{{|\varphi _m(0)\rangle: m = 1\ldots N}\right\}
\]
remains frozen in time. 
\item
\label{cond2N}
There exists a collection  $\left\{\psi '_n (t)\right\}_{n = 1\ldots N}$ which provides unitary bases for the domain spaces
\[
F(t)=\mbox{span}\left\{{|\psi _n'(t)\rangle: n = 1\ldots N}\right\}.
\]
and, moreover, all the vectors satisfy the one-particle Shr\"odinger equation, i.e.
\begin{equation}
\label{kNdschr}
i\hbar \frac{d}{dt}\langle\psi _n'| = -\langle\psi _n'| H.
\end{equation}
\end{enumerate}
When both conditions \ref{cond1N} and \ref{cond2N} are satisfied, then solutions of the mesoscopic equation (\ref{MesoSchr}) admit representation in the form
\begin{equation}
\label{kNdsol}
K(t) = \sum [A'(t)]_{m n} |\varphi _m (0)\rangle\langle\psi _n'(t)|,
\end{equation}
where 
\begin{equation}
\label{Atime1}
A'(t)=\sqrt{A'(0)A'(0)^*}
\exp\left(\frac{i}{\hbar}\int_{0}^{t}B^2(t')(A'(0)A'(0)^*)^{-1}dt'\right).
\end{equation}
\label{theor2} 
\end{th}

Naturally, Proposition \ref{prop1} is a special case of Theorem \ref{theor2}. We introduced it beforehand not only because it is interesting in itself, but also because it provides a smooth introduction into the internal logic of the problem. As before, all strictly analytic issues, such as \emph{existence} and \emph{regularity} of solutions of the nonlinear problem (\ref{MesoSchr}) rest on the corresponding properties of the \emph{linear} Schr\"odinger equation (\ref{kNdschr}). Needless to say, vast literature is available in relation to the latter theme. 

I would also like to point out that Theorem \ref{theor2} conforms with Theorem \ref{theor1}. Both theorems expose a rather unobvious fact that the mesoscopic evolution is factored through the Schr\"odinger mechanics. Indeed, the $W$ factor of Theorem \ref{theor1} encodes the Schr\"odinger evolution. Of course, Theorem  \ref{theor2} is not a mere corollary, and its proof required additional arguments, while also relying on the former theorem.  Finally, let me emphasize that no conclusions have been drawn here as to the infinite-dimensional case, i.e. the case when (the nontrivial part of) the domain of $K(t)$ cannot be encapsulated in a finite-dimensional space $F(t)$. There are many other questions of interest not even attempted here, particularly those pertaining to the important case when the Schr\"odinger operator $H$ itself depends on the dynamic variable $K$. The closing section explains the nature and significance of such a feedback.  

\section{The broader context}
\label{sec4}

We will devote these concluding remarks to sketching the broader perspective in which the results of this article ought to be seen.  First, I would like to point out that these results do not generalize to other types of operator equations. Indeed, the unique type of nonlinearity in the mesoscopic Schr\"odinger equation plays a crucial role in the proofs. Specifically, it allows separation of the radial and the unitary part in the polar decomposition of the dynamic variable. Secondly, I would like to point out that the result is important in view of the physical interpretation of the MeM. Indeed, it shows that only the Schr\"odinger particles, i.e. electrons, can participate in the mesoscopic transport. This is not guarantied {\em a priori}, e.g. some other type of, say, a nonlinear wave could appear in place of the Schr\"odinger waves, which possibility is hereby {\em a posteriori} excluded. I will now provide a short synopsis of the physical interpretation of the solutions according to the theory that has been put forward in \cite{MeM}.   

Electromagnetic phenomena in vacuum are described by the classical Maxwell equations. These equations are modified by the so-called material constants or even by introduction of nonlinearities as modelers adapt them to describe propagation of the electromagnetic wave in various materials. Such an approach is usually sufficient when the model is meant to reflect what happens at the macroscopic scale.  We know from experiment that in low temperatures some materials feedback to the electromagnetic field in a more profound way. Namely, at the nano-scale the spacial distribution of the magnetic field depends on the quantum picture of the electronic structure of the material. This fact is of particular importance  in the context of high-temperature superconductivity and the Quantum Hall Effects. A problem arises, how to describe the inter-relation of the ambient magnetic field and the electronic structure. The particular form of this interrelation has far reaching consequences as regards the resulting galvanomagnetic characteristics of the material, e.g. cf. reference \cite{Falko} in which some aspects of this problem are analyzed assuming random distribution of the magnetic field.   

The Mesoscopic Mechanics postulates that the distribution of the magnetic field assumes a particular form depending on the quantum characteristic of the material. Specifically, let us focus attention on an idealized planar electronic system characterized by the single-particle Hamiltonian $H$. Imagine this system being exposed to a perpendicular magnetic field with magnetic induction $B$. As a result of the interaction between the ambient magnetic field and the electronic structure, the magnetic flux will get distributed over the surface area nonuniformly. Here, depending on the properties of the system, the single-particle Hamiltonian $H$ may or may not depend on the magnetic field. It is a basic precept of the MeM that even if the Hamiltonian does not depend on the magnetic field, the flux distribution may still be nonuniform. Specifically, the MeM postulates that the distribution of flux is determined in a certain way by an operator $K$ of the type considered in this article.  Namely, let $\Phi$ be the total magnetic flux through the surface, and let $\Upsilon$ stand for the coherent state 
\[
\Upsilon = \sum_{\mbox{\small{filled states}}}  |\psi_n\rangle,
\] 
which in this way accounts for the actual distribution of electron states. The \emph{simplest} postulate of the MeM is that (with an appropriate normalization) the distribution of the magnetic flux is approximately
\[
(x,t)\rightarrow \Phi|K(t)\Upsilon |^2(x).
\] 
Of course, the interpretation is probabilistic. Moreover, the evolution of the system is described via $K$ by the mesoscopic Schr\"{o}dinger equation (\ref{MesoSchr}). This equation of motion is determined by the total Hamiltonian $\Xi$, cf. equation (\ref{Ksi}). The total Hamiltonian accounts for the single-electron portion of the energy, as well as the inter-electron phase correlation energy. This latter energy is enclosed in the determinant (or {\em entropy}) term of $\Xi$, which is switched on with an application of the magnetic field. I emphasize that we are looking at a new type of interaction of the magnetic field with the Fermi sea, independent and separate from the phenomenon of formation of Landau states. Electrons respond collectively since they are bound together by the energy of phase correlation. Naturally, the specific features of this phenomenon strongly depend on the energy-band structure of the material. Further information can be found in \cite{MeM}. 

Finally, let me point out that the total Hamiltonian $\Xi$ is related to the following functional (whose arguments are functions)
\begin{equation}
L_A(\psi )= \int |\nabla _A\psi|^2 + B^2\int\ln(|\psi |^2).
\label{functional}
\end{equation}
Naturally, the logarithmic integral mimics the entropy term.  Some properties of a particular realization of this latter functional, especially as regards magnetic-vortex type critical points, are described in an earlier article \cite{sowa3}.  Last but not least, the MeM has a field-theoretic counterpart, the Nonlinear Maxwell Theory, cf. 
\cite{sowa4}, which embraces (\ref{functional}) as one of its central objects, and provides models for many low-temperature phenomena. 


\end{document}